**Cognitive Bias Detection Using Advanced Prompt Engineering**[1]

[Frederic Lemieux][2], Aisha Behr[3], [Clara Kellermann-Bryant][4], [Zaki Mohammed][5]

**Abstract**

Cognitive biases, systematic deviations from rationality in judgment, pose significant challenges in generating objective content. This paper introduces a novel approach for real-time cognitive bias detection in user-generated text using large language models (LLMs) and advanced prompt engineering techniques. The proposed system analyzes textual data to identify common cognitive biases such as confirmation bias, circular reasoning, and hidden assumption. By designing tailored prompts, the system effectively leverages LLMs' capabilities to both recognize and mitigate these biases, improving the quality of human-generated content (e.g., news, media, reports). Experimental results demonstrate the high accuracy of our approach in identifying cognitive biases, offering a valuable tool for enhancing content objectivity and reducing the risks of biased decision-making.

**Introduction**

Cognitive biases are systematic patterns of deviation from rational judgment, affecting decision-making processes across various domains, including media, policy-making, and legal reasoning. With the rapid expansion of artificial intelligence (AI) applications, large language models (LLMs) have demonstrated significant potential in processing and evaluating vast amounts of textual information. However, existing research has largely focused on mitigating biases within AI-generated outputs rather than leveraging AI to detect biases in human-generated content. This gap presents a critical challenge in ensuring transparency and fairness in AI-assisted decision-making. This study explores the application of structured prompt engineering as a novel approach to improving LLM accuracy in detecting cognitive biases. Prior research has identified key cognitive biases such as circular reasoning, false causality, and confirmation bias in various forms of communication. While existing bias detection models rely on conventional natural language processing (NLP) techniques, they often struggle with contextual accuracy and tend to produce high false-positive rates.

[1] The authors want to thank the *Initiative on Pedagogical Uses of AI (IPAI)* at Georgetown University and CGI Federal for supporting this research.
[2] Professor of the Practice and Faculty Director, Georgetown University.
[3] Director Consulting Services, AI & Machine Learning, CGI Federal.
[4] Graduate Student and Research Assistant, Georgetown University.
[5] Graduate Student and Research Assistant, Georgetown University.

Our research introduces an optimized framework for designing structured prompts that enhance LLMs' ability to distinguish between biased and neutral statements more effectively. By conducting a comparative analysis against baseline models, this study evaluates the impact of structured prompt engineering in reducing misclassification and improving bias detection accuracy. The findings from this research contribute to the broader discourse on AI's role in cognitive bias detection by demonstrating that well-crafted prompting strategies can significantly enhance model performance. This study also highlights the need for continued refinement in AI-driven bias detection methodologies, emphasizing the importance of human annotation, multilingual adaptation, and real-world application testing.

**Literature Review**
Research on biases in artificial intelligence (AI) systems has focused on developing tools and strategies to detect and mitigate biases within automated systems, algorithms, and data. Early studies, such as those by Liang and Acuna (2020), introduced psychological frameworks to detect biases in AI, notably through the identification of gendered perceptions in sentiment analysis and word embeddings. These efforts highlighted how cognitive biases inherent in human behavior often propagate through AI outputs. Alelyani (2021) extended this by examining cognitive biases in machine learning datasets, revealing that human biases in data collection, such as those found in Amazon's hiring systems, lead to algorithmic discrimination. Her research called for greater transparency in datasets and model explainability to effectively mitigate these biases. The work of Raza et al. (2024) introduces Nbias, a framework designed to detect and mitigate biases in textual data across domains like social media, healthcare, and job hiring. By using a transformer-based token classification model, the framework identifies biased words/phrases, achieving accuracy improvements of 1% to 8% over baseline models while promoting the fair and ethical use of data. Rastogi et al. (2022) proposed a time-based de-biasing strategy for human-AI collaboration, enhancing decision-making by addressing biases like anchoring during interaction. These studies illustrate the complexity of bias in AI systems and the need for real-time intervention strategies.

This literature review primarily focuses on the detection of bias through natural language processing (NLP) techniques and the role of large language models (LLMs) in identifying cognitive biases in text. However, it is important to acknowledge that other approaches to bias detection and mitigation exist. Algorithmic auditing, as discussed by Raji et al. (2020), offers a framework for systematically evaluating biases embedded in AI models, particularly in high-stakes applications such as hiring and criminal justice. Fairness-aware machine learning models, such as those explored by Dwork et al. (2012), incorporate fairness constraints to mitigate biases in decision-making processes. Additionally, debiasing techniques in computer vision and facial recognition systems, studied by Buolamwini and Gebru (2018), have demonstrated the prevalence of racial and gender biases in automated image classification systems. These alternative approaches highlight the multifaceted nature of bias detection and mitigation, underscoring the need for interdisciplinary solutions to address cognitive biases across AI applications.

Incorporating foundational psychological research, the seminal work of Tversky et al. (1982) on heuristics and biases provides a strong theoretical underpinning for understanding how cognitive biases influence decision-making. Their studies on anchoring, availability, and representativeness heuristics laid the groundwork for contemporary discussions on bias detection in AI. Gigerenzer and Gaissmaier (2011) further examined the interplay between heuristics and rational decision-making, which has direct implications for how AI models detect and interpret cognitive biases in user-generated text. By integrating these psychological insights, AI systems can be better designed to recognize and correct for cognitive distortions that arise in human communication.

From an ethical AI perspective, scholars such as Timnit Gebru and Joy Buolamwini (2018) have highlighted the risks of bias amplification in AI models, particularly in facial recognition and natural language processing applications. Their research underscores the need for algorithmic transparency and accountability, which aligns with this study's emphasis on real-time bias detection. Binns (2018) explored fairness in machine learning and how different philosophical approaches to fairness—such as distributive and procedural justice, impact bias mitigation strategies. These ethical considerations are crucial when developing AI systems designed to identify and address cognitive biases in human-generated content.

In industry-specific applications, biases in AI decision-making have been widely studied across various domains. In healthcare, Feehan et al. (2021) investigated how cognitive biases influence clinical decision-making, leading to diagnostic errors and disparities in patient treatment. AI tools capable of detecting these biases in medical records and decision-support systems can enhance objectivity and patient outcomes. In finance, Barocas et al. (2023) examined the role of algorithmic bias in credit scoring and loan approvals, where historical biases in financial data disproportionately affect marginalized groups. Similarly, in the legal field, Almasoud and Idowu (2024) explored the impact of bias in predictive policing and risk assessment algorithms, demonstrating how biased training data can perpetuate systemic inequalities. By applying bias detection techniques across these industries, AI can contribute to fairer decision-making processes and reduced disparities in automated decision systems.

Despite the progress in developing tools to mitigate biases, much of the research has focused on bias detection in AI-generated outputs, with limited exploration of the potential for AI to detect biases in human-generated content. Emerging literature highlights the promising role of AI in identifying cognitive biases in human communication. Atreides and Kelley (2024) demonstrated that LLMs could reliably detect 188 cognitive biases in text, such as confirmation bias and circular reasoning, outpacing human identification capabilities. Similarly, Parsapoor (2023) showed that AI could detect cognitive impairments like early Alzheimer's disease through speech and language analysis, illustrating AI's potential to identify subtle patterns in human communication that may indicate cognitive biases. Lee et al. (2024) further emphasized AI's ability to detect biases by incorporating human gaze patterns into explainable AI systems, improving both decision-making speed and accuracy.

While these studies have proven the feasibility of AI tools for bias detection, gaps remain in addressing the full spectrum of real-world applications. Few studies have systematically compared AI's performance across diverse textual contexts. Atreides and Kelley (2024) identified limitations in detecting biases related to memory and ambiguity, while Zhu et al. (2024) noted the need for improved evaluation of AI-generated outputs, as human cognitive biases influence how AI performance is perceived. This article addresses these gaps by proposing a systematic framework for training LLMs to recognize cognitive biases accurately, integrating prompt engineering and real-world applications to improve objectivity, transparency, and decision-making. By bridging theoretical research with practical use cases, the study offers a scalable solution for detecting cognitive biases in human communication.

The development of LLMs capable of detecting cognitive biases in textual content has profound implications across various fields. In healthcare, cognitive biases such as confirmation bias and anchoring bias can influence clinical judgments and treatment decisions (Lee et al., 2024). In the legal field, biases like framing effects and stereotyping impact judgments, legal drafting, and policymaking (Alelyani, 2021). Finally, biases in corporate decision-making, such as overconfidence and status quo bias, can result in flawed risk assessments and missed opportunities (Atreides & Kelley, 2024). The ability of LLMs to detect and mitigate these biases presents an opportunity to enhance decision-making, fairness, and transparency across diverse professional domains.

Previous studies have made significant strides in identifying and mitigating cognitive biases within AI systems and machine learning algorithms, but the majority of research has focused on detecting biases within AI-generated outputs or datasets. Fewer studies have explored the potential for AI, specifically large language models (LLMs), to detect cognitive biases in human-generated content. Existing work has largely been limited to identifying biases in controlled environments or through post-hoc analysis, without offering robust methodologies for improving detection accuracy through optimized AI interactions. This research contributes to the field by introducing a systematic framework for cognitive bias detection in user-generated text using advanced prompt engineering techniques. Unlike previous efforts that rely on standard NLP methodologies, this study emphasizes the rigor of prompt engineering as a fundamental approach to improving bias detection accuracy. By designing structured and contextually adaptive prompts, the research ensures that LLMs effectively identify a wide range of cognitive biases, such as confirmation bias, straw man fallacy, and circular reasoning, while minimizing hallucination and misclassification errors. Furthermore, this work goes beyond prior studies by demonstrating the scalability and adaptability of prompt engineering for cognitive bias detection across various domains. The framework is tested in practical contexts such as news reporting, social media, healthcare, legal professions, and corporate decision-making, illustrating how well-crafted prompts enhance bias recognition in diverse linguistic settings. In addressing these gaps, the research not only advances the methodological robustness of AI-driven bias detection but also lays the groundwork for future studies that refine LLM interactions for ethical AI applications. This contribution underscores the importance of structured AI prompting as a key factor in improving bias detection transparency and fairness in decision-making processes.

**Methodology**

*Operationalization of concepts*

The study examined the ability of a Large Language Model (LLM) and prompt engineering to detect common cognitive biases. For this study, we focused on six common cognitive biases and fallacies: Straw Man, False Causality, Circular Reasoning, Mirror Imaging, Confirmation Bias, and Hidden Assumptions. These biases were selected based on their prevalence in decision-making and discourse. Each of these biases follows a specific pattern of reasoning, which allows us to construct structured prompts that mirror their logical sequence, aiding in their detection Straw man refers to misrepresenting or oversimplifying an opponent's argument to make it easier to refute. This can lead to a misunderstanding of the actual position and a failure to address the real issues. False causality is a logical fallacy where a causal relationship is incorrectly assumed or established between two events or variables. This fallacy occurs when it is assumed that because one event follows another, the first event must be the cause of the second, without sufficient evidence to support this causal connection. Circular Reasoning refers to using a conclusion to support the assumption that was necessary to reach that conclusion. This creates a loop in reasoning where the evidence and conclusion are the same. Mirror imaging implies Assuming that other actors (states, organizations, individuals) will act or react in the same way as one's own country or group, based on one's own values, priorities, and decision-making processes. This can lead to misunderstandings of intentions and capabilities. Confirmation bias leads to selectively searching for, interpreting, and recalling information that confirms pre-existing beliefs, while ignoring or dismissing contradictory evidence. Finally, a hidden assumption, also known as an implicit assumption, is an unstated premise or belief that underlies an argument, decision, or belief system. It is not explicitly expressed or acknowledged but is necessary for the argument or decision to be coherent or valid (Watson et al., 2024; Pohl, 2004).

*Data Collection & Preprocessing*

For the purpose of this study, open-source data was collected from diverse sources, including X, Reddit, Medium, ResearchGate, the University of Michigan Library, the Library of Congress, The Economic History of India-Oxford Academic, Marines.com, Whitehouse.gov, the Israel US Embassy, Europol, and DHS. Sample texts derived from these sources provided a wide range of text types for evaluating the model's performance. The use of open-source data from various sources ensured a broad representation of text types and potential biases. In addition, the team selected varying lengths of text to ensure our model was able to handle this kind of variation.

To ensure a diverse and balanced dataset, we categorized data sources into three levels of rigor: low, medium, and high. Low-rigor sources, such as politicians' speeches, social media posts, and personal blogs, undergo little to no review and often reflect personal opinions. Medium-rigor sources, including newspaper articles and podcast transcripts, may have some editorial oversight but still contain biases. High-rigor sources, such as think tank reports, academic papers, and policy briefs, are subjected to thorough peer or institutional reviews, ensuring greater credibility and trustworthiness. Additionally, human annotators deliberately sought out biases from opposing perspectives on controversial issues, such as gun control, abortion, and climate change. By incorporating biases from both sides of debates, we ensured a balanced representation of different viewpoints, making the dataset more comprehensive.

Next, we processed the collected data by converting each text file into a LangChain Document object. To facilitate efficient processing by the LLM, we employed the Recursive Character Text Splitter technique to segment each document into smaller, more manageable chunks. This chunking process ensured that the input sequences were compatible with the LLM's context window while preserving the semantic integrity of the text. To evaluate the effectiveness of our approach, we designed a robust annotation pipeline. First, the processed text was incorporated into a carefully crafted prompt template and input to the LLM. The resulting LLM-generated output, along with the original text sample, was then passed to our human annotation system, powered by Argilla. Argilla, a user-friendly platform for data annotation and analysis in NLP, facilitated the efficient review and labeling of our data. This human-in-the-loop approach allowed annotators of varying expertise to readily interact with the text and LLM outputs, providing valuable feedback for evaluating our approach and ensuring the accurate identification of cognitive biases.

To enhance objectivity in content evaluation and annotation, this study employed a rigorous evaluation process. Human annotators were trained to identify the selected cognitive biases using the logical pattern followed by each cognitive bias and fallacy, providing a necessary framework to assess the LLM's performance. Their task involved reviewing the LLM-generated responses and evaluating their accuracy in detecting or dismissing the presence of bias. This human evaluation served as a crucial validation step, ensuring the reliability and effectiveness of the LLM responses. The human annotators' expertise in identifying nuanced instances of cognitive biases provided valuable feedback for refining the system and enhancing its overall performance. Following the human evaluation and annotation of the dataset, we proceeded to the analysis phase. This phase aimed to rigorously assess the performance of our model in accurately detecting the presence or absence of various cognitive biases within the text. Specifically, we evaluated the model's ability to identify each predefined cognitive bias component, such as confirmation bias, straw man fallacy, and circular reasoning. This analysis allowed us to determine the accuracy of our model in recognizing these biases and provided crucial insights into its overall effectiveness in detecting and flagging potentially biased language.

*Prompt Engineering Methodology*

The core contribution of this research is the rigor of prompt engineering for cognitive bias detection. Unlike traditional NLP-based classification models, our approach employs structured and optimized prompts designed to mimic cognitive bias reasoning patterns. Cognitive biases often follow systematic sequences of thought, making them detectable through engineered prompts that guide the LLM in identifying logical inconsistencies. Our methodology leveraged prompt engineering to effectively steer the LLM towards accurate bias detection. Contextual cues were embedded to help the LLM recognize subtle forms of bias while mitigating hallucinations. By structuring prompts to align with cognitive bias patterns, we enhanced detection accuracy beyond conventional classification models.  A structured prompt template was constructed, consisting of two key components: 1) A set of explicit directives outlining the specific type of bias to be identified and 2) the text input to be analyzed. This structured approach allowed us to systematically control the LLM output space, effectively constraining the LLM's generation process, shifting the probability distribution towards the desired outcome and increasing the likelihood of accurate and relevant bias detection.

*Experimental Design*

For this study, we selected the Mixtral 7x8B instruct model as our foundation and leveraged the Langchain framework to design and implement our prompt templates. This choice was motivated by several key strengths of Mixtral 7x8b. Firstly, its instruction variant is specifically fine-tuned to follow instructions and perform tasks as defined in the prompt, making it ideal for complex prompt engineering. Secondly, despite its relatively smaller size compared to some larger LLMs, Mixtral 7x8b demonstrates impressive performance across various benchmarks, including those related to reasoning and understanding natural language. This balance of size and performance makes it a suitable choice for real-time applications where computational efficiency is crucial. Finally, its open-weight nature allows for greater transparency and customization, enabling us to further refine the model for our specific bias detection task. Utilizing Langchain for prompt construction streamlined the process of managing and optimizing complex prompt structures, facilitating experimentation with different prompting strategies.

To evaluate the effectiveness of our prompt engineering approach, we benchmarked our model against two baseline models. Our first baseline utilized the Mixtral 7x8B instruct model, but forewent our specialized prompt template. Instead, we used a simple prompt instructing the model to detect the specific cognitive bias under consideration without the set of explicit directives given in our prompt engineering strategy. This allowed us to isolate the impact of our prompt engineering. Our second baseline utilized the Llama 3 70B instruct model, also with a basic prompt. We included Llama 3 70B, a significantly larger language model, to assess the influence of model scale on bias detection performance. Comparing our approach with both a mid-sized model like Mixtral and a large-scale model like Llama 3 provides valuable insights into the interplay between model size, prompt engineering, and bias detection accuracy.

*Evaluation Metrics & Human Annotation*

The human annotation phase was crucial for evaluating the performance of our cognitive bias detection system and was divided into two distinct phases.

Phase 1: This phase focused solely on assessing the accuracy of the LLM's responses when employing our novel prompt engineering technique. We meticulously examined whether the LLM, guided by our prompts, could effectively identify instances of cognitive biases in the text data. This initial evaluation allowed us to ensure the LLM was properly equipped to analyze the text for biases.

Phase 2: In this phase, we broadened our evaluation to compare the accuracy of our model against two established baseline models. Human annotators independently reviewed the responses from all three models and compared them against their own expert judgment. This comparative analysis provided valuable insights into the relative strengths and weaknesses of each approach, further validating the effectiveness of our proposed system. To ensure a comprehensive evaluation, we employed a multi-faceted approach to determining the accuracy of the LLM responses generated using our prompt engineering technique. Accuracy was established under the following conditions:

*Agreement on Bias Detection*: Both the LLM and the human annotator identified the presence of the same cognitive bias in the text.
*Agreement on No Bias*: Both the LLM and the human annotator agreed that no cognitive bias was present in the text.
*Agreement on Ambiguity*: Both the LLM and the human annotator concluded that it was unclear whether a cognitive bias existed in the text.

Any deviation from these conditions, where the human annotator disagreed with the LLM's assessment, resulted in the LLM response being classified as incorrect. This rigorous evaluation process ensured a robust and reliable assessment of our system's performance in detecting cognitive biases. During the first phase of human annotation, labelers assessed multiple factors, including whether the sample text exhibited any bias, the accuracy of the model's responses in detecting the presence or absence of bias, and, in cases of incorrect model outputs, they supplied a correct response based on the logical framework established by the university, as illustrated in *Figure 1: Phase 1 Human Annotation System*. This phase was essential in aligning the model with the intended system objectives, ensuring a strong foundation for precise and unbiased results.

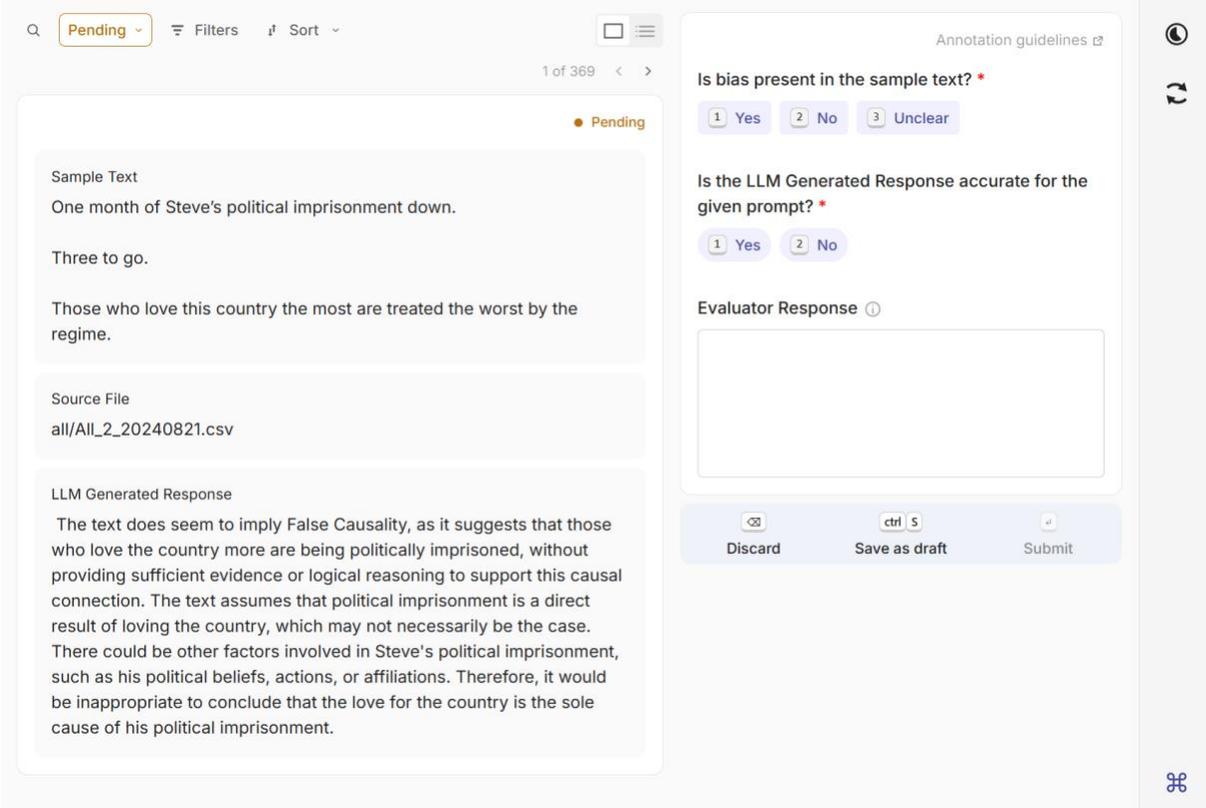

*Figure 1: Phase 1 Human Annotation System*

During the second phase, the human laborers evaluated our model's responses against two baseline models as shown in *Figure 2: Phase 2 Human Annotation System*. This comparison enabled us to assess our model not just in terms of intent alignment, but also in relation to the responses generated by other models that did not incorporate our design approach. This comprehensive assessment provided valuable insights into the strengths of our model, highlighting its unique capabilities and the impact of our design choices on its performance.

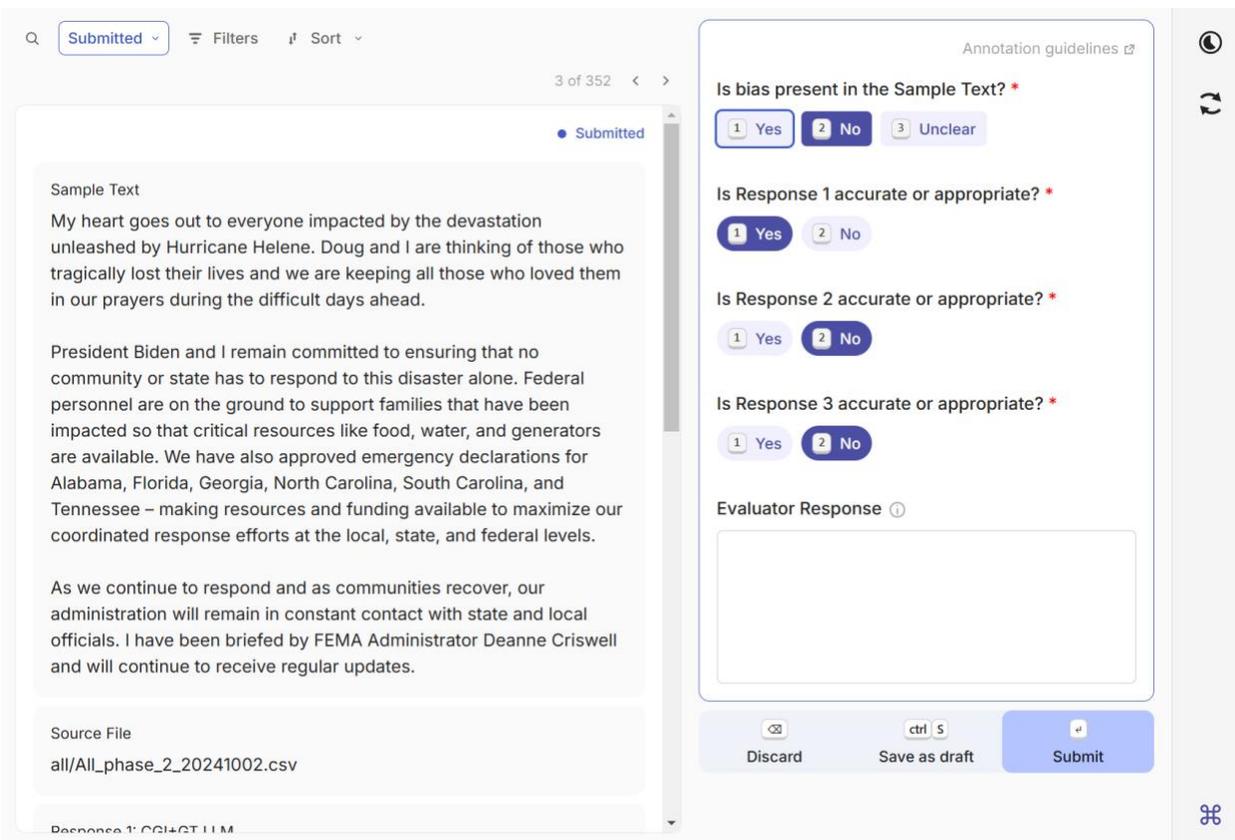

*Figure 2: Phase 2 Human Annotation System*

**Results**

This study was conducted in two phases. Phase 1 focused on testing and refining our prompt-engineered model for detecting cognitive biases, while Phase 2 compared our optimized model against two baseline models: Mixtral 7x8B (without our prompt strategy) and Llama 3 70B (without our prompt strategy). The results of both phases demonstrate the effectiveness of structured prompt engineering in improving bias detection accuracy.

Phase 1
In the first phase, a dataset of 4,321 texts containing various cognitive biases was selected and analyzed by our model. These texts were curated to ensure equal representation across all six cognitive biases. The results indicate that our model was able to determine whether bias was present with an accuracy of at least 96% (*Table 1*). The model achieved perfect accuracy in detecting circular reasoning and performed exceptionally well in identifying confirmation bias, false causality, hidden assumptions, mirror imaging, and straw man fallacy. Circular reasoning had the highest accuracy, suggesting that the model effectively recognizes the logical loop inherent in this bias. False causality, however, had a slightly lower accuracy, which could be attributed to the challenge of distinguishing causation from correlation in textual data. Overall, the results from Phase 1 demonstrate that our model can reliably detect cognitive biases in user-generated content with a high degree of accuracy.

| Bias | Correct | Incorrect | Accuracy |
|---|---|---|---|
| Circular Reasoning | 442 | 0 | 100.00 |
| Confirmation Bias | 721 | 12 | 98.36 |
| False Causality | 610 | 25 | 96.06 |
| Hidden Assumption | 725 | 1 | 99.86 |
| Mirror Imaging | 1144 | 7 | 99.39 |
| Straw Man Fallacy | 619 | 15 | 97.63 |

*Table 1: Accuracy of Our model with Phase 1 dataset*

Phase 2

The second phase sought to compare our optimized model against two baseline models to assess the impact of structured prompt engineering on bias detection. An additional dataset of 2,160 texts was evaluated using our model that incorporated our prompt strategy, the Mixtral 7x8B baseline model, and the Llama 3 70B baseline model. Each model was tested on 305 distinct sample texts per cognitive bias to ensure a fair comparison. Our model outperformed both baseline models in all categories, achieving nearly perfect accuracy across all six biases. The results revealed that structured prompt engineering plays a crucial role in enhancing model performance. While our model consistently identified biases with high accuracy, the baseline models struggled, particularly with more nuanced biases such as false causality and hidden assumptions. *Figure 3 presents* the distribution of the Bias Detection Types vs the Cognitive Bias Types. It is important to note that while the majority of the collected samples are a "no" as the Bias Detection Type for each Cognitive Bias Type, if the model evaluated the response as "no", then that is still considered an accurate response.

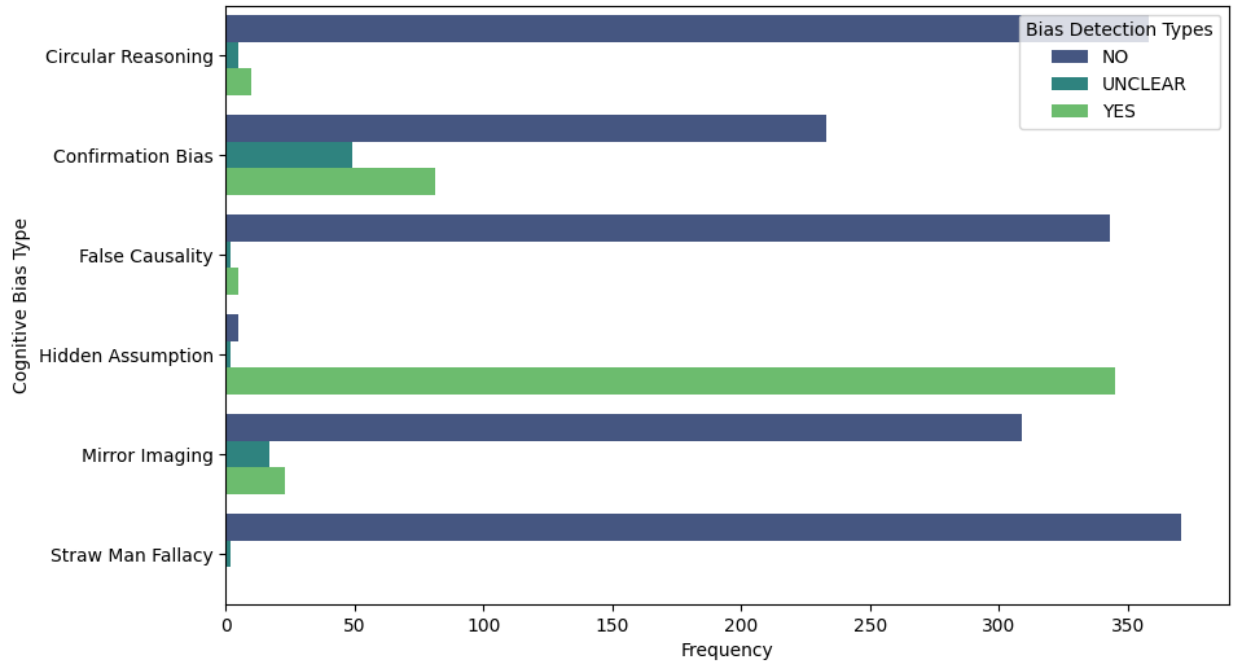

*Figure 3: Distribution of Bias Detection Types by Cognitive Bias*

For Circular reasoning, our model correctly identified if a sample text had bias, did not have bias, or was deemed unclear for 373/373 samples, as shown in *Table 2*. For Confirmation bias, 360/363 samples were correctly identified with our model. For False causality, 350/350 samples were correctly identified. For Hidden assumption, 352/352 samples were correctly identified. For Mirror imaging, 349/349 samples were correctly identified. For Straw man fallacy, 373/373 samples were correctly identified.

| Bias | Correct | Incorrect | Accuracy |
| --- | --- | --- | --- |
| Circular Reasoning | 373 | 0 | 100.00 |
| Confirmation Bias | 360 | 3 | 99.17 |
| False Causality | 350 | 0 | 100.00 |
| Hidden Assumption | 353 | 0 | 100.00 |
| Mirror Imaging | 349 | 0 | 100.00 |

| Straw Man Fallacy | 373 | 0 | 100.00 |

Table 2: Accuracy of Our model in Phase 2

The comparative analysis highlights the significant performance gap between models with and without a set of well crafted directives in the structured prompt (see *Figure 2*). For example, in detecting circular reasoning, our model correctly identified bias in all 373 samples, whereas the Mixtral 7x8B baseline model identified only 209 samples correctly, and the Llama 3 70B baseline model correctly identified just 150 samples. A similar trend was observed across other biases, with the accuracy of the baseline models dropping sharply compared to our structured approach. The Llama 3 70B model, despite its larger size, underperformed in comparison to our optimized Mixtral 7x8B model. This suggests that model size alone does not compensate for the lack of structured prompting and that effective prompt engineering is a more significant determinant of bias detection performance than sheer model scale.

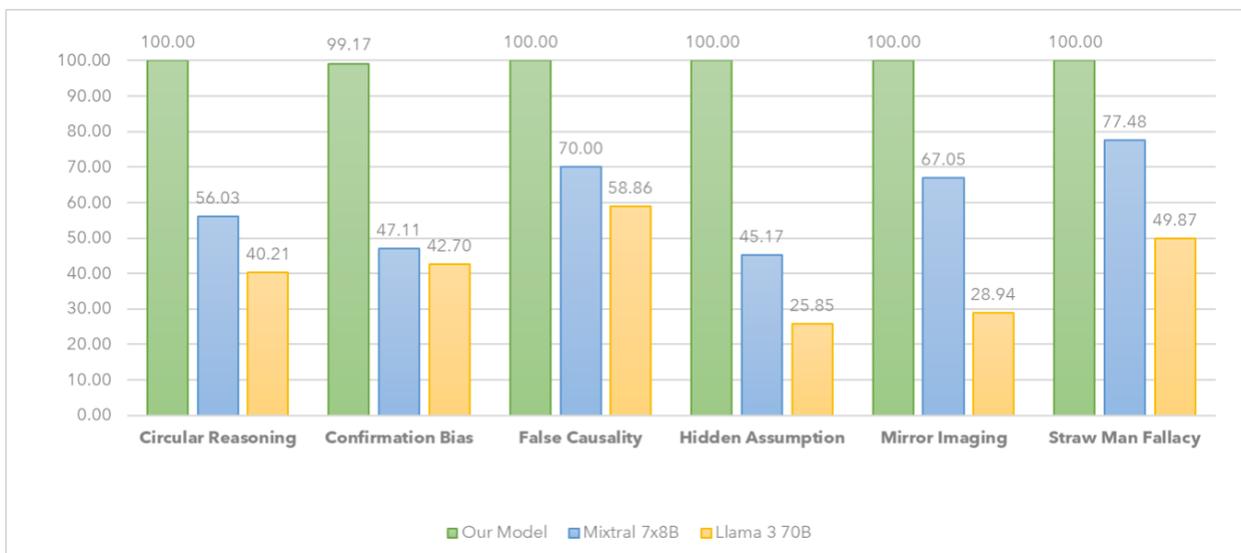

*Figure 2: Our Model vs. Baseline Models*

**Analysis and Discussion**

The results of this study confirm that well crafted directives within the structured prompt engineering strategy substantially improves the accuracy of LLM-based cognitive bias detection. By guiding the model through structured reasoning patterns, our approach enabled more precise differentiation between biased and neutral statements. One of the key advantages of our methodology was its ability to reduce false positives by refining the detection process through human annotators and ensuring that biases were identified with contextual accuracy. The comparison with baseline models further reinforces the effectiveness of structured prompting, demonstrating that even a mid-sized model with optimized prompts can outperform a much larger model with basic prompting techniques.

Another key finding is that different biases vary in their ease of detection. Circular reasoning and hidden assumptions were consistently identified with high accuracy, likely because their logical structures are more explicit and easily recognized. In contrast, biases such as false causality presented a greater challenge, suggesting that distinguishing causation from correlation in text remains an area for future research. The strong performance of our model indicates that structured prompt engineering not only enhances bias detection but also contributes to improving transparency in AI-driven content evaluation. The ability to systematically surface and explain biases within user-generated text has profound implications for media analysis, policy research, and automated decision-support systems.

These findings align with prior research discussed in the literature review, particularly the work of Atreides and Kelley (2024), which demonstrated that automated systems can detect a broad range of cognitive biases. Similarly, Xie et al. (2024) emphasized the potential of multi-agent systems to detect subtle biases that might evade traditional NLP methods. However, unlike previous studies that focused primarily on detecting biases within AI-generated outputs, our approach is uniquely designed to analyze and detect biases within human-generated content, bridging a crucial gap in existing research. Furthermore, while prior research, such as that of Zhu et al. (2024), highlighted the cognitive biases that humans exhibit toward AI-generated content, our findings suggest that AI, when properly structured through engineered prompts, can effectively serve as a tool for enhancing objectivity in bias detection. Also, our results are particularly strong when compared to prior studies in the field, such as the work by Raza et al. (2024), who reported F1-scores of 88.4%, 90.6%, and 91.8% for their model in detecting Social Media Bias, Health-related Bias, and Job Hiring Bias, respectively. While Raza's model demonstrated solid performance, our approach with structured prompt engineering showed superior results in cognitive bias detection across all biases tested, surpassing the F1-scores achieved by Raza's Nbias model. This suggests that, while Raza's framework works well in certain applications, our model benefits from optimized prompting strategies that enhance both the precision and recall of bias detection.

Despite the strengths of our approach, there are several limitations to consider. First, while our model demonstrated high accuracy in detecting explicit biases, more nuanced and context-dependent biases may still pose challenges. Biases that rely heavily on domain-specific knowledge or require deep contextual understanding may require further prompt refinement or hybrid approaches that incorporate additional linguistic or factual verification mechanisms. Additionally, our evaluation focused primarily on English-language text, meaning that the generalizability of our findings to other languages and cultural contexts remains an open question. Future research should explore multilingual bias detection and assess whether structured prompting techniques can be effectively adapted to different linguistic structures.

Another limitation is the reliance on human annotation as a benchmark for evaluating bias detection performance. While human annotators were trained using a standardized framework, cognitive biases are inherently subjective, and interpretations of bias may vary among individuals. This introduces a level of subjectivity that, while mitigated through inter-annotator agreement measures, may still impact overall evaluations. Future studies could explore ways to refine human annotation methods, such as incorporating expert review panels or using adversarial testing to assess model robustness. Finally, while our model significantly outperformed baseline approaches, ongoing improvements in LLMs suggest that larger-scale models with enhanced reasoning capabilities could further refine bias detection. Future work should investigate whether fine-tuning specialized models for cognitive bias detection, rather than relying solely on prompt engineering, could yield additional improvements in accuracy and reliability.

**Conclusion**

This study establishes that prompt engineering significantly enhances the accuracy of LLM-based cognitive bias detection. The results from both phases demonstrate that structured prompting leads to a substantial performance increase compared to baseline models that lack this optimization. The findings also indicate that model size alone is not the primary determinant of bias detection accuracy; rather, well-crafted prompting strategies play a more critical role. Despite the strengths of our approach, there are several limitations that warrant further exploration. The reliance on English-language data constrains the generalizability of the model across linguistic and cultural contexts, emphasizing the need for multilingual adaptations in future studies. Additionally, while our model effectively detected explicit biases, more complex biases requiring deeper contextual analysis may necessitate alternative approaches, such as fine-tuning specialized models for domain-specific bias detection. Moreover, the subjective nature of human annotation introduces an inherent challenge in validating bias detection accuracy, necessitating ongoing refinements in human-in-the-loop AI evaluation methods. Future research should focus on expanding the dataset to include more complex and domain-specific biases, refining prompts to further reduce false positives, and exploring real-world deployment scenarios. Additionally, investigating the applicability of structured prompting techniques in multilingual settings and domain-specific fields would further enhance the robustness and generalizability of AI-driven bias detection systems. The insights gained from this study can serve as a foundation for improving AI-driven content moderation, misinformation detection, and automated bias identification across multiple sectors.